\documentclass[aps,prd,superscriptaddress,a4paper,nofootinbib]{revtex4}

\usepackage{latexsym}
\usepackage{amsmath}
\usepackage{amssymb}
\usepackage{ulem}
\usepackage{fancyhdr}
\usepackage{natbib}
\usepackage{url}
\usepackage{graphicx}
\usepackage{grffile}
\usepackage[vcentering,dvips]{geometry}
\usepackage{sidecap}

\begin{document}

\title{Gravitational waves in a de Sitter universe}

\author{Nigel~T. Bishop}
\affiliation{
Department of Mathematics, Rhodes University, Grahamstown 6140, South
Africa
}

\begin{abstract}

The construction of exact linearized solutions to the Einstein equations within the
Bondi-Sachs formalism is extended to the case of linearization about de Sitter
spacetime. The gravitational wave field measured by distant observers is constructed,
leading to a determination of the energy measured by such observers. It is found that
gravitational wave energy conservation does not normally apply to inertial observers,
but that it can be formulated for a class of accelerated observers, i.e. with
worldlines that are timelike but not geodesic.

\end{abstract}

\maketitle

\section{Introduction}

The standard formulation of gravitational wave (GW) theory assumes an asymptotically
flat spacetime (for example, see~\cite{Bondi62,Penrose:1965}). However, astrophysical
evidence has emerged over the last two decades that the Universe is undergoing an
accelerated expansion which is well described by the $\Lambda$CDM model. Thus it is
important to investigate GW properties in such a spacetime, and further it is in a
way urgent to do so since direct GW detection by facilities such as LIGO is regarded
as imminent. In order to be able to model this problem, the first step is to decide on
a background spacetime to be used. For ease of analysis it should, in some way, be
analytically simple, yet for any results to be astrophysically relevant it should
also be realistic. These two requirements are somewhat contradictory, and here we
use de Sitter spacetime. Although the spacetime is unrealistic in that it has no
matter content, it does include the key feature of an accelerated expansion. The
key advantage for the present study of de Sitter spacetime is its simplicity.

In a series of recent papers, Ashtekar {\it et
al.}~\cite{Ashtekar-2015,Ashtekar-2015-1,Ashtekar-2015-2,Ashtekar-2015-3} have
investigated GWs in de Sitter spacetime. The focus of their work has been on
the asymptotic structure and correct mathematical formulation of GWs. In addition,
they have tackled the astrophysical issues of whether a de Sitter background would affect
generation or detection of GWs, and concluded that the effect is negligible. There
has also been recent related work by Date and Hoque~\cite{Date-2015}.
In this paper we study similar matters, although from a different perspective.

Exact linearized solutions about a background spacetime within the Bondi-Sachs
formalism have also received recent attention~\cite{Cedeno-2015-1,
Cedeno-2015-2}. The simplest background to consider is Minkowski,
and~\cite{Bishop-2005b} constructed such solutions for the purpose of providing a
test-bed for numerical relativity codes using the Bondi-Sachs formalism. Subsequently,
the approach has been used to construct solutions about Schwarzschild~\cite{Madler2013},
to investigate quasi-normal modes~\cite{Bishop:2009}, to find the spacetime geometry
generated by binaries in circular orbit~\cite{Bishop:2011,Cedeno-2015-2}, and to
generalize previous results to the case of sourcing by arbitrary matter
fields~\cite{Cedeno-2015-1}. It is therefore natural to extend this approach further
by seeking to linearize about de Sitter spacetime, particularly in the light of the
more general issues raised above.

In order to linearize about a given spacetime, its metric in Bondi-Sachs form must
first be found. This is surprisingly difficult. It is of course easy to do so for
both Minkowski and Schwarzschild. It can be done for Kerr~\cite{Bishop:2006} but
the result involves elliptic integrals; the simplest cosmological model is
Einstein-de Sitter, but the Bondi-Sachs form of its metric has to be constructed
numerically.
It is therefore somewhat noteworthy that the Bondi-Sachs metric for de Sitter spacetime
is very simple, and further that the solutions for linearized perturbations take
a simple polynomial form. As such, one may hope to be able to use the solutions to
gain physical insight into the perturbations, i.e. the GWs, in de Sitter spacetime.

The analytic solutions obtained are valid throughout the Bondi-Sachs domain external
to the source. Conformal compactification is not introduced, and so issues involving the
treatment of ${\mathcal J}^+$ are avoided. We do, however, consider the leading order term
in an expression, and this is equivalent to taking an asymptotic limit.
The GW analysis is from the physical viewpoint of the geodesic deviation that would
be measured by an observer. In order for the usual concept of conservation of GW
energy to apply, this quantity would be expected to have an asymptotic fall-off of
$1/r$. Perhaps surprisingly, it is found that this fall-off does not apply to inertial
observers (i.e., on a timelike geodesic), but only to a class of accelerated observers.
Thus we believe that the study of the solutions presented here can provide useful
physical insight, which may be helpful towards a proper understanding of GW energy in
de Sitter spacetime.

As well as investigating GWs from the viewpoint of a detector, we also investigate the
effect of a de Sitter background on the generation of GWs. This is achieved by a
straightforward adaptation of the procedure about Minkowski~\cite{Bishop:2011}. As
reported in other work, we also find that the consequence of the change in background
is normally completely negligible. Certainly this is the case for direct detection of
GWs, and it is difficult to conceive of any measurable astrophysical process that would
be indirectly affected.

In Section~\ref{s-dS2BS} the coordinate transformation from usual de Sitter coordinates
to Bondi-Sachs form is determined and implemented. Then the linearized Einstein
equations are constructed and solved in Section~\ref{s-lEe}, with the properties of
the associated GWs developed in Section~\ref{s-GWs}. GWs from an equal mass binary
in de Sitter spacetime are calculated (Section~\ref{s-GW=m}), and the paper ends
with the Conclusion in Section~\ref{s-Conc}.  Throughout the paper, extensive use is
made of computer algebra, and the associated scripts are described in
Appendix~\ref{a-cas}; further the scripts are available in the online supplement.

\section{Transformation of de Sitter spacetime from comoving to Bondi-Sachs coordinates }
\label{s-dS2BS}
De Sitter spacetime, in standard comoving coordinates $(t,\rho,\theta,\phi)$, is represented by the metric
\begin{equation}
ds^2=-dt^2+A^2 e^{2\alpha t}\left(d\rho^2+\rho^2 d\Omega^2\right)\,,
\label{e-dS}
\end{equation}
where $A$ is a constant, $\alpha=\sqrt{\Lambda/3}$ ($\Lambda$ is the cosmological
constant), and $d\Omega^2=d\theta^2+\sin^2\theta d\phi^2$. Now make the coordinate
transformation
\begin{equation}
(t,\rho)\rightarrow (u,r): u=-\frac{\log(A\rho\alpha\exp(t\alpha)+1)-t\alpha}{\alpha},\;\;
r=A\rho e^{\alpha t}\,,
\end{equation}
with inverse
\begin{equation}
t=\frac{\log(1+r\alpha)+u\alpha}{\alpha},\;\;\rho=\frac{r\exp(-u\alpha)}{A(1+r\alpha)}\,.
\label{e-BS2dS}
\end{equation}
The metric~(\ref{e-dS}) is transformed to
\begin{equation}
ds^2=-du^2(1-r^2\alpha^2)-2du\,dr+r^2d\Omega^2\,,
\label{e-dSinBS}
\end{equation}
which is in Bondi-Sachs form since it is clear that $r$ is both a null and a surface
area coordinate. Note that each null cone $u=u_0=$ constant has a maximum radius in
de Sitter coordinates  $\rho_{\mbox{max}}=\exp(-u_0\alpha)/(A\alpha)$ (from
Eq.~(\ref{e-BS2dS})), but $r$ is unbounded.

We will also need the transformation for the 4-velocity when the motion is purely radial.
$V^a$ in Bondi-Sachs coordinates
\begin{equation}
V^a_{[BS]}=\left(V^0,\frac{1}{2V^0}-\frac{V^0(1-r^2\alpha^2)}{2},0,0\right)
\label{e-VaBS}
\end{equation}
becomes in de Sitter coordinates
\begin{equation}
V^a_{[dS]}=\left(\frac{1}{2V^0(1+r\alpha)}+\frac{V^0(1+r\alpha)}{2},
\frac{1}{2V^0(1+r\alpha)^2}-\frac{V^0}{2},0,0\right)\,.
\label{e-VadS}
\end{equation}
In particular, we note that a ``natural'' de Sitter observer for which
$V^a_{[dS]}=(1,0,0,0)$ satisfies
\begin{equation}
V^a_{[BS]}=\left(\frac{1}{1+r\alpha},r\alpha,0,0\right)\,.
\label{e-VNdS}
\end{equation}

\section{The linearized Einstein equations and their solution}
\label{s-lEe}

The formalism for expressing Einstein's equations as an evolution system based
on characteristic, or null-cone, coordinates is based on work originally due to Bondi
{\it et al.}~\cite{Bondi1960,Bondi62} for axisymmetry, and extended to the general case
by Sachs~\cite{Sachs62}. The formalism is covered
in the review by Winicour~\cite{Winicour05}, and the conventions used here
are as~\cite{Bishop:2014}. We start with coordinates based upon a
family of outgoing null hypersurfaces.
Let $u$ label these hypersurfaces, $x^A$ $(A=2,3)$ label
the null rays, and $r$ be a surface area coordinate. In the resulting
$x^\alpha=(u,r,x^A)$ coordinates, the metric takes the Bondi-Sachs
form
\begin{equation}
 ds^2  =  -\left(e^{2\beta}\left(1 + \frac{W}{ r}\right) -r^2h_{AB}U^AU^B\right)du^2
        - 2e^{2\beta}dudr -2r^2 h_{AB}U^Bdudx^A 
        +  r^2h_{AB}dx^Adx^B,
\label{eq:bmet}
\end{equation}
where $h^{AB}h_{BC}=\delta^A_C$ and
$\det(h_{AB})=\det(q_{AB})$, with $q_{AB}$ a metric representing a unit
2-sphere; for the computer algebra calculations presented later it is necessary
to have specific angular coordinates, and for that purpose stereographic
coordinates are used with $x^A=(q,p)$ and
\begin{equation}
q_{AB}dx^A dx^B=\frac{4}{(1+q^2+p^2)^2}\left(dq^2+dp^2\right).
\end{equation}
$W$ is related to the usual Bondi-Sachs
variable $V$ by $V=r+W$. It is convenient to
describe angular quantities and derivatives by means of complex numbers using the
spin-weighted formalism and the eth ($\eth$) calculus~\cite{Bishop96,Bishop97b,Gomez97}.
To this end, $q_{AB}$ is represented by a complex dyad $q_A$
with, for example, $q_A=(1,i)2/(1+q^2+p^2)$ in stereographic coordinates. Then
$h_{AB}$ can be represented by its dyad component $J=h_{AB}q^Aq^B/2$.
We also introduce the field $U=U^Aq_A$. (The field $K=\sqrt{1+J \bar{J}}$ is
needed in the nonlinear case but not here because linearization implies $K=1$).

The procedure for linearizing about de Sitter spacetime is very similar to that used
for linearization about other fixed backgrounds. So here we omit some detail, and just
outline the key steps highlighting the differences between the current case and
that of Minkowski spacetime as described in~\cite{Bishop-2005b}.
The Bondi-Sachs metric quantities
\begin{equation}
J, \beta, U, w 
\end{equation}
(where $w=r^3\alpha^2+W$) are regarded as being small (${\mathcal O}(\epsilon)$), and
all terms in the Einstein equations of order ${\mathcal O}(\epsilon^2)$ are set to zero.
Most calculations are in vacuum, but if matter is present it will be assumed that its
density $\rho={\mathcal O}(\epsilon)$. An important difference
between this paper and previous work is that here the Einstein equations include the
cosmological constant, so that the equations to be evaluated are
\begin{equation}
R_{ab}-3\alpha^2 g_{ab}=4\pi(2T_{ab}-T g_{ab}).
\end{equation}
We find:
\begin{equation}
R_{11}: \;\; \frac{4}{r}\beta_{,r}=8 \pi T_{11}
\label{e-b}
\end{equation}
\begin{equation}
q^A R_{1A}: \;\; \frac{1}{2r} \left(
4 \eth \beta - 2 r \eth \beta_{,r} + r \bar{\eth} J_{,r}
+r^3 U_{,rr} +4 r^2 U_{,r} \right) = 8 \pi q^A T_{1A}
\label{e-rq}
\end{equation}
\begin{equation}
h^{AB} R_{AB}: \;\;
(4(1-3r^2\alpha^2)-2\eth \bar{\eth}) \beta +\frac{1}{2}(\bar{\eth}^2 J + \eth^2\bar{J})
+\frac{1}{2r^2}(r^4\eth\bar{U}+r^4\bar{\eth}U)_{,r} -2 w_{,r}
=8 \pi (h^{AB}T_{AB}-r^2 T)
\label{e-rw}
\end{equation}
\begin{equation}
q^A q^B R_{AB}: \;\;
  -2\eth^2\beta + (r^2 \eth U)_{,r} - 2(r - 2\alpha^2 r^3) J_{,r}
 - \left( 1 - \alpha^2 r^2 \right) r^2 J_{,rr} 
  +2 r (rJ)_{,ur}= 8 \pi q^A q^B T_{AB}.
\label{e-ev}
\end{equation}
The remaining Einstein equations are needed only in the vacuum case, and are
\begin{align}
R_{00}:\;\;&
 \frac{1}{2r^3} \bigg( r(r-\alpha^2 r^3) w_{,rr}+\eth\bar{\eth} w 
  +2(r-\alpha^2 r^3) \eth\bar{\eth} \beta
  + \alpha^2 r^4 (\eth \bar{U} + \bar{\eth}U) \nonumber \\
  &+ 12 r^3 \alpha^2 (1-r^2\alpha^2)\beta
-4 r (r-\alpha^2 r^3) \beta_{,u} - r^3 (\eth \bar{U} + \bar{\eth}U)_{,u}+2 r w_{,u}
 \bigg) = 0
 \label{e-R00}
\end{align}
\begin{equation}
R_{01}:\;\;
 \frac{1}{4r^2} \bigg(2 r w_{,rr}+4 \eth\bar{\eth}\beta +24\beta\alpha^2 r^2
         -(r^2\eth\bar{U}+r^2\bar{\eth}U)_{,r}\bigg)=0
\end{equation}
\begin{align}
q^A R_{0A}:\;\;&
 \frac{1}{4r^2} \bigg( 2r \eth w_{,r}-2 \eth w+ 2 r^2(r-r^3\alpha^2)(4 U_{,r}
         + r U_{,rr})+4 r^2 U +r^2(\eth\bar{\eth}U-\eth^2\bar{U})\nonumber \\
         &+2 r^2 \bar{\eth}J_{,u}-2 r^4  U_{,ur}-4 r^2 \eth\beta_{,u}
         \bigg)=0.
\label{e-R0A}
\end{align}

The solution to the above equations is constructed as described in~\cite{Bishop-2005b}.
We make a standard ``separation of variables'' ansatz
\begin{align}
\beta &=\beta_0(r) \Re(\exp(i\omega u)) Z_{\ell m},\;
w=w_0(r) \Re(\exp(i\omega u)) Z_{\ell m},\nonumber \\
U &=U_0(r) \Re(\exp(i\omega u))\,{}_1 Z_{\ell m},\;
J=J_0(r) \Re(\exp(i\omega u)) \,{}_2 Z_{\ell m},
\end{align}
where the $Z_{\ell m}$ are real spherical harmonics, and the ${}_s Z_{\ell m}$ are
their spin-weighted extensions (see~\cite{Bishop-2005b,Zlochower03}). An important
technical difference between this ansatz and that of~\cite{Bishop-2005b} is that
here we use ${}_s Z_{\ell m}$ as basis functions, instead of $\eth^s Z_{\ell m}$. The relation between the two sets of basis functions is
\begin{equation}
\eth^s Z_{\ell m} =\sqrt{\frac{(\ell-s)!}{(\ell+s)!}}\, {}_s Z_{\ell m}\,.
\end{equation}
In the vacuum
case, Eqs.~(\ref{e-b}) to (\ref{e-R0A}) reduce to a system of ordinary differential
equations in $\beta_0(r),w_0(r),U_0(r),J_0(r)$.  It is remarkable that, following the
same procedure as~\cite{Bishop-2005b} for the case of linearization about a
Minkowski background, these equations can be solved exactly, and that for the
outgoing wave the solutions are simple polynomials in $1/r$. Eq.~(\ref{e-b}) shows
that $\beta_0=$ constant, then Eqs.~(\ref{e-rq}) and (\ref{e-ev}) may be combined
to give (in the case $\ell=2$)
\begin{equation}
x^2(x^2-\alpha^2)\frac{d^2J_2(x)}{dx^2}+2x(2x^2+i\omega x+\alpha^2)\frac{dJ_2(x)}{dx}
-2(2x^2+i\omega x+\alpha^2)J_2(x)=0\,,
\label{e-meq}
\end{equation} 
where $x=1/r$ and $J_2(x)=d^2 J(x)/dx^2$. Eq.~(\ref{e-meq}) has a solution of the form
$J_2(x)=C_3 f_3(x)+C_4 f_4(x)$, but one of the functions, say $f_4(x)$ is of the form
$\exp(2i\omega r)$ representing incoming radiation so normally we set $C_4=0$. Integrating
$J_2(x)$ to obtain $J_0(r)$ introduces two further integration constants $C_1,C_2$.
The procedure leading to Eq.~(\ref{e-meq}) also leads to an expression for $U_0(r)$
in terms of $J_0(r)$ and so can now be evaluated. Next Eq.~(\ref{e-rw}) is integrated
to give $w_0(r)$ in terms of an additional integration constant $C_5$. The constraint
equations Eqs.~(\ref{e-R00}) to (\ref{e-R0A}) impose two conditions on the constants
of integration, and are used to express $C_2,C_5$ in terms of $\beta_0,C_1,C_3$.

We find in the leading order $\ell=2$ case:
\begin{align}
\beta_0(r) &=\mbox{constant}\nonumber \\
U_0(r)&= \sqrt{6}\left(\frac{-24 i\omega\beta_0+3C_1(3\alpha^2+\omega^2)
-C_3\omega^2(4\alpha^2+\omega^2)}{36} +\frac{2\beta_0}{r}
+\frac{C_1}{2r^2}+\frac{iC_3\omega}{3r^3}+\frac{C_3}{4r^4}\right)\nonumber \\
w_0(r)&= -2\beta_0\alpha^2 r^3 +\frac{r^2}{6}\left(24i\omega\beta_0
 -3C_1(\omega^2+3\alpha^2)+\omega^2 C_3(\omega^2+3\alpha^2)\right)\nonumber \\
 &+\frac{r}{3}\left(-6\beta_0+3i\omega C_1 -C_3 i \omega(\omega^2+4\alpha^2)\right)
 -C_3(\omega^2+\alpha^2)+\frac{i\omega C_3}{r}+\frac{C_3}{2r^2}
 \nonumber \\
J_0(r)&= \sqrt{6}\left( \frac{24\beta_0- i (4\omega\alpha^2+\omega^3) C_3+3i\omega C_1}{18}+\frac{C_1}{2r}-\frac{C_3}{6r^3}\right)\,.
\label{e-UwJ}
\end{align}
This solution has been checked by substituting it into all 10 vacuum Einstein equations,
and confirming that $R_{ab}-3\alpha^2 g_{ab}=0$.

In the case of linearization about Minkowski, the modes with $\ell>2$ also have solutions
that are polynomial in $1/r$ but of higher order. The highest power of $1/r$ appears in
$U_0(r)$ and is $1/r^{\ell+2}$. While it has not been explicitly checked, it can
reasonably be expected that the same situation would apply to linearization about
de Sitter.

\section{Gravitational waves}
\label{s-GWs}

In the asymptotically flat case, GWs are described in terms of the wave strain in
the TT gauge ($h_++ih_\times$), the gravitational news (${\mathcal N}$) or the
Newman-Penrose quantity $\psi_4$. These descriptors are related by time derivatives
(and constant factors due to the use of different conventions in the original
definitions), specifically
\begin{equation}
r\psi_4=2\partial_u\bar{\mathcal N}=r\partial^2_u (h_+-ih_\times)\,.
\label{e-psi4NH}
\end{equation}
In order to decide which descriptor is most convenient here, the issue of gauge
freedom needs to be considered, which is whether results obtained would be affected
if a small change (${\mathcal O}(\epsilon)$) were to be made to the coordinates. In
such a case calculations are more difficult because the coordinates have to be
specified to satisfy geometrical conditions to ${\mathcal O}(\epsilon)$. Now,
both $h_++ih_\times$ and ${\mathcal N}$ are gauge dependent quantities, but $\psi_4$
is not. The Newman-Penrose quantity $\psi_4=-C_{abcd}n^a\bar{m}^bn^c\bar{m}^d$ is
defined in terms of the Weyl tensor $C_{abcd}$ and a null tetrad (specified below).
It is tensorially a scalar, but gauge freedom can appear in the specification of
the null tetrad. However, in the background de Sitter metric $C_{abcd}=0$ so $\psi_4$,
and indeed all the $\psi_i$, are gauge independent quantities.

Suppose that an observer has 4-velocity $V^a$ given
by Eq.~(\ref{e-VaBS}). Then the null tetrad ($n^a,\ell^a,m^a$) must be chosen so that
\begin{equation}
\sqrt{2} V^a=n^a+\ell^a,
\label{e-Vnl}
\end{equation}
and for the unperturbed ($\epsilon=0$) spacetime is
\begin{equation}
n^a=\left(\sqrt{2}V^0,-\frac{V^0(1-r^2\alpha^2)}{\sqrt{2}},0,0\right),\;\;
\ell^a=\left(0,\frac{1}{V^0\sqrt{2}},0,0\right),\;\;
m^a=\left(0,0,\frac{q^A}{\sqrt{2}r}\right)\,,
\end{equation}
where $q^A$ is the dyad on the unit sphere. Evaluating $\psi_4$, we obtain
\begin{align}
\psi_4&=(V^0)^2 {}_{-2}Z_{2m}\times \nonumber  \\
&\Re\left[\exp(i\omega u)C_3 \frac{\sqrt{6}}{12}\left(
-\frac{2\omega^4+8\omega^2\alpha^2+3\alpha^4}{r}
+i\frac{4\omega^3+10\omega\alpha^2}{r^2}
+6\frac{\omega^2+\alpha^2}{r^3}
-6i\frac{\omega}{r^4}-\frac{3}{r^5}\right)\right] \,.
\label{e-psi4}
\end{align}
Note further that the forms of $\psi_0,\cdots,\psi_3$ have been checked, and they have
the expected fall-off behaviour $C_3r^{-5},\cdots,C_3r^{-2}$. It is also important
to note that the formula for $\psi_4$ involves only $C_3$, and not $\beta_0,C_1$.
This is expected as the same applies in the asymptotically flat case to all the GW
descriptors. Thus, as in the Minkowski case, the constants $\beta_0,C_1$ may
be regarded as representing gauge freedoms.

The physical relevance of $\psi_4$ needs to be addressed, i.e. the relationship between
$\psi_4$ and geodesic deviation (which is what is actually measurable by a detector such
as LIGO). In a vacuum $\Lambda=0$ spacetime $R_{ab}=0$ so that $C_{abcd}=R_{abcd}$,
and the relevant geodesic deviation $R_{abcd}V^am^bV^cm^d$ can be expressed, using
Eq.~(\ref{e-Vnl}), as $C_{abcd}(n^a+\ell^a)m^b(n^c+\ell^c)m^d/2$, which to leading
order in asymptotic fall-off is $-\bar{\psi}_4/2$. In the de Sitter case,
$R_{ab}=\Lambda g_{ab}\ne 0$ so the preceeding argument cannot be applied. However,
the result is still true since
\begin{align}
\left(C_{abcd}-R_{abcd}\right)&(n^a+\ell^a)m^b(n^c+\ell^c)m^d\nonumber \\
=&\Lambda\left(-g_{a[c}g_{d]b}+g_{b[c}g_{d]a}+\frac{4 g_{a[c}g_{d]b}}{3}\right)
(n^a+\ell^a)m^b(n^c+\ell^c)m^d\,,
\label{e-C-R}
\end{align}
and every term on the right hand side involves two inner products of null tetrad vectors,
of which at least one inner product must vanish. (The only non-zero inner product
in Eq.~(\ref{e-C-R}) is $g_{ab}n^a\ell^b=-1$). Thus there is an equivalence between
the gauge invariant quantity $\psi_4$ and geodesic deviation.

However, for a physical interpretation of detectable gravitational waves, we need to do rather  more than just evaluate $\psi_4$. Assuming that the detector is in free-fall and so
follows a timelike geodesic, allowance needs to be made for changes to the position and
velocity of the detector, and further all results should be in terms of the detector's
proper time $\tau$ rather than the Bondi-Sachs time coordinate $u$. Let $\partial_\tau
\psi_4$ mean the rate of change of $\psi_4$ as observed by the detector, then
\begin{equation}
\partial_\tau \psi_4
=V^0\partial_u \psi_4+V^1\partial_r \psi_4+V^0\partial_{V^0}\psi_4\,.
\label{e-dpsi4dtau}
\end{equation}
The first term in Eq.~(\ref{e-dpsi4dtau}) would appear in a calculation about
Minkowski and represents the red-shift factor; and the second term is unimportant
because $\partial_r \psi_4={\mathcal O}(1/r^2)$ and so the term does not make a
leading order contribution. The third term however is highly significant. From the
geodesic equation we find $\partial_\tau V^0=-\alpha^2 r (V^0)^2$, so it is non-zero
only in the de Sitter case; futher the multiplication by $r$ means that the leading
order part of the expression is affected. The expression found is rather long, and
we present only the part to ${\mathcal O}(1/r)$
\begin{equation}
\partial_\tau \psi_4=\frac{5(V^0)^3\sqrt{6}C_3\alpha^2
(8\alpha^2\omega^2+2\omega^4+3\alpha^4)}{24}
-i\frac{(V^0)^3\sqrt{6}C_3\omega(33\alpha^4+2\omega^4+20\alpha^2\omega^2)}{12r}\,.
\label{e-dpsi4_r-1}
\end{equation}

In the $\Lambda=0$ asymptotically flat case, the well-known ``News = mass loss''
theorem~\cite{Bondi62} applies. A key precursor to the result is that the magnitude
of the GWs decays as $r^{-1}$ so that
\begin{equation}
\int_{S}|{\mathcal N}|^2\,,
\end{equation}
(where $S$ is a spherical shell $u=r=$constant) is independent of $r$. For de Sitter
spacetime, our objective is to investigate conditions under which energy conservation
in the above sense applies. Using Eqs.~(\ref{e-dpsi4_r-1}) and (\ref{e-psi4NH})
the total energy crossing a 2-surface $r=u=$ constant according to observers with
4-velocity $V^a$ is
\begin{align}
\int_{S}|{\mathcal N}|^2=\frac{(V^0)^2 C_3^2}{96\omega^4}\bigg[&
r^2\alpha^4(225\alpha^{8}+1200\alpha^6\omega^2+1900\alpha^4\omega^4
+800\alpha^2\omega^6+100\omega^8)\nonumber \\
&+\omega^2(16\omega^8+320\omega^6\alpha^2+2128\omega^4\alpha^4+5280\omega^2\alpha^6
+4356\alpha^8)\bigg]\,.
\end{align}
The above can be made independent of $r$ by appropriate choice of $V^0$. To
leading order in $\alpha$,
\begin{align}
V^a=\bigg(&D\left(1-\frac{10}{\omega^2}\alpha^2
+\frac{668-25r^2\omega^2}{8\omega^4}\alpha^4\right),\nonumber \\
&\frac{1-D^2}{2D}
+\frac{10+D^2(10+r^2\omega^2)}{2\omega^2 D}\alpha^2
+\frac{132+25r^2\omega^2+D^2(668+55r^2\omega^2)}{16\omega^4 D}\alpha^4
,0,0\bigg)\,,
\label{e-VaD0}
\end{align}
where $D$ is a constant.

\subsection{Discussion}
We now discuss the implications for GW energy conservation of the results above.
For observers in free-fall, energy conservation applies provided
Eq.~(\ref{e-VaD0}) is satisfied. In order to make the implications more concrete, we
consider the example where the constant $D$ is fixed by the condition $V^0\rightarrow 1$
as $r\rightarrow 0$; physically, this corresponds to observers close to the source being
at rest relative to the source. In this case, we find
\begin{equation}
V^a=\left(1-25\frac{r^2\alpha^4}{\omega^2},\frac{r^2\alpha^2}{2}\left(1+50\frac{\alpha^2}
            {\omega^2}\right),0,0\right)+{\mathcal O}(\alpha^6)\,.
\end{equation}
Thus, while it is possible to find a set of observers for whom energy conservation holds,
their velocities depend on both position and wave frequency. If the wave
frequency changes, as
happens during an inspiral, then the observer's velocity would need to be adjusted, i.e.
the observer would need to be accelerated.

Abandoning the idea that the observers for whom energy conservation holds should be
freely falling, we can, for example, set $V^0=1$ everywhere (so that physically
observers near the source are at rest relative to the source). In order that such
observers have $\partial_\tau V^0=0$, they must experience an acceleration
$-r\alpha^2+r^3\alpha^4/2$.

We note that the ``natural'' de Sitter observers are freely falling and have $V^a$
given by Eq.~(\ref{e-VNdS}): they do not constitute a set of observers for whom energy
conservation applies.

An important question is whether a positive cosmological constant will
affect the interpretation of observations expected from detectors such as
LIGO. The effects are ignorable provided, from Eq.~(\ref{e-dpsi4_r-1}),
$r\alpha^2\ll\omega$. This is indeed the case, since the lowest $\omega$ can be is a few
Hz, $\alpha\approx$ 1/(5 Gpc), and the largest expected value of $r$ is of order
1 Gpc for a binary black hole merger. A similar conclusion was reported by
Ashtekar {\it et al.}~\cite{Ashtekar-2015}.

\section{Gravitational waves from an equal mass binary}
\label{s-GW=m}

A procedure for calculating the gravitational field, linearized about Minkowski,
for two equal mass $M$ objects in circular orbit radius $r_0$ was described
in~\cite{Bishop:2011}, and that result has recently been generalized~\cite{Cedeno-2015-2}.
The calculations about de Sitter and Minkowski proceed in the same way,
and here we just provide an outline.

The objects are modeled as point particles, and so the matter density $\rho(u,r,x^A)$
is expressed in terms of $\delta$-functions. It is then straightforward to decompose
$\rho$ into spherical harmonic components, i.e. $\rho=\Sigma \rho_{\ell m} Z_{\ell m}$.
For the case $\ell=2$, $\rho_{21}=\rho_{2,-1}=0$ and $\rho_{20}$ is constant in time
and thus is not the source of any radiation. Only $\rho_{22}$ and $\rho_{2,-2}$
are relevant. Turning now to the metric, there are two separate solutions, valid
in $r<r_0$ and $r>r_0$ respectively, with each solution having its own set
of integration constants. The exterior solution has 3 constants, and the interior
solution has only 1 free constant (with the others fixed by the condition that
spacetime must be regular at the origin). These constants are fixed by imposing 4
conditions at the interface $r=r_0$: continuity of $J$ and $U$; and jump conditions
on $\beta,w$ that follow from integrating the $\delta$-functions in the right hand 
sides of Eqs.~(\ref{e-b}) and (\ref{e-rw}) across $r=r_0$.

The result obtained for the metric in the region $r>r_0$ is
\begin{align}
\beta&=\Re(\beta_0 \exp(i\omega u))Z_{22}+\Re(-i\beta_0 \exp(i\omega u))Z_{2,-2}
\nonumber\\
w&=\Re(w_0 \exp(i\omega u))Z_{22}+\Re(-w_0 \exp(i\omega u))Z_{2,-2}
\nonumber\\
J&=\Re(J_0 \exp(i\omega u))\,{}_2Z_{22}+\Re(-iJ_0 \exp(i\omega u))\,{}_2Z_{2,-2}
\nonumber\\
U&=\Re(U_0 \exp(i\omega u))\,{}_1Z_{22}+\Re(-iU_0 \exp(i\omega u))\,{}_1Z_{2,-2}
\,,
\end{align}
where $\omega$ is the wave frequency which is twice the orbital frequency; and where
$w_0,J_0,U_0$ are given by Eqs.~(\ref{e-UwJ}), with the integration constants taking
the values
\begin{align}
\beta_0&=\frac{M}{r_0}\sqrt{15\pi}\,,\nonumber \\
C_1&=-\frac{2M}{3}\sqrt{15\pi}+\frac{M}{3}i\omega r_0\sqrt{15\pi}
+\frac{2M}{15}\omega^2 r_0^2\sqrt{15\pi}-\frac{8M}{15}\alpha^2 r_0^2\sqrt{15\pi}
+{\mathcal O}(r_0^3,\alpha^4)\,,\nonumber \\
C_3&=\frac{4M}{5}r_0^2\sqrt{15\pi}-\frac{4M}{5}i\omega r_0^3\sqrt{15\pi}
-\frac{64M}{105}\omega^2r_0^4\sqrt{15\pi}-\frac{24M}{35}\alpha^2r_0^4\sqrt{15\pi}
+{\mathcal O}(r_0^5,\alpha^4)\,.
\label{e-orbit}
\end{align}
As discussed earlier the GW field is determined by the value of $C_3$. From its
expression in Eq.~(\ref{e-orbit}) it is
clear that the effect of a de Sitter rather than Minkowski background is insignificant
if $\alpha^2r_0^2 \ll 1$. It is difficult to envisage any astrophysical scenario in
which that would not be the case.

\section{Conclusion}
\label{s-Conc}

Within the Bondi-Sachs formalism, exact solutions linearized about a de Sitter
background have been constructed, and these have been used to investigate some
properties of GWs in that spacetime. It was found that the most convenient GW
descriptor to use is $\psi_4$ because of its gauge invariance, and it was also
shown that $\psi_4$ is physically relevant because it is directly related to
the geodesic deviation measured by an observer.

The paper investigated the gravitational wave energy determined by various
observers. In particular, it was found to be difficult to define observers in
free fall for which energy conservation of the GWs would apply. Such observers
would need to have velocities that depend on position (which is not surprising)
but also on the wave frequency. On the other hand, it was found to be straightforward
to define accelerated observers for which energy conservation applies.

The paper determined the gravitational field around an equal mass binary, showing
to leading order the addtional terms introduced on using a de Sitter background.

Changing from a Minkowski to a de Sitter background introduced modifications to
formulas for the generation, propagation  and detection of GWs. In realistic
astrophysical scenarios these modifications are so small as to be completely
ignorable. However, the modifications may be significant for the mathematical
understanding of the concept of GW energy in a de Sitter spacetime, since in
particular the result that inertial observers are not appropriate for describing GW
energy is somewhat counter-intuitive. The development of a proper theory of mass loss
and GW energy is a matter for future work.

\acknowledgments
This work was supported by the National Research Foundation, South Africa. I thank
the Perimiter Institute for Theoretical Physics, Canada, and Universitat de les Illes
Balears, Spain, for hospitality support while conducting this work. I am grateful
to Sascha Husa, Julien Larena, Luis Lehner and Denis Pollney for discussions.

\appendix
\section{Computer algebra scripts}
\label{a-cas}

The Maple scripts used in this work are given in the online supplement, and their
purposes are summarized here.
\begin{itemize}
\item The script {\bf dS.map} evaluates the coordinate transformation from de Sitter
to Bondi-Sachs coordinates to obtain Eqs. (\ref{e-dSinBS}) and (\ref{e-VadS}).
\item The script {\bf lin.map} uses ProcsRules.map and gamma.out, and evaluates
Eqs.~(\ref{e-b}) through (\ref{e-R0A}). It also checks that the solution
Eq.~(\ref{e-UwJ}) satisfies all 10 Einstein equations.
\item The script {\bf masterEqn.map} evaluates Eq.~(\ref{e-meq}).
\item The script {\bf l2.map} solves Eqs.~(\ref{e-b}) through (\ref{e-R0A}) to obtain
the solution Eq.~(\ref{e-UwJ}).
\item The script {\bf weyl.map} uses ProcsRules.map and gamma.out, and evaluates
the Weyl tensor $C_{abcd}$ to find $\psi_i$ and in particular $\psi_4$ in
Eq.~(\ref{e-psi4}). It further evaluates Eqs.~(\ref{e-dpsi4_r-1}) through (\ref{e-VaD0}).
\item The script {\bf GWfromBinary.map} evaluates Eq.~(\ref{e-orbit}).
\item The files {\bf ProcsRules.map} and {\bf gamma.out} are auxiliary files used by other
scripts as stated above.
\end{itemize}
\bibliography{paper,aeireferences}

\end{document}